\documentclass[twocolumn,showpacs,aps]{revtex4}
\usepackage[dvips]{graphicx}
\usepackage{bm}
\usepackage{mathrsfs}

\begin{document}
\date{\today}

\title{Geometric Flow appearing in Conservation Law in Classical and Quantum Mechanics}
\author{Naohisa Ogawa
\footnote{ogawanao@hus.ac.jp}}
\affiliation{Hokkaido university of Sciences, Sapporo 006-8585 Japan}

\begin{abstract}
The appearance of a geometric flow in the conservation law of particle number in classical particle 
diffusion and in the conservation law of probability in quantum mechanics is discussed in the geometrical environment 
of a  two-dimensional curved surface with thickness $\epsilon$ embedded in $R_3$.
 In such a system with a small thickness $\epsilon$, the usual two-dimensional conservation law does not 
 hold and we find an anomaly by using the equation $\partial \rho / \partial t  + \nabla_i^{(2)}J^i \neq 0$,
where  $\rho$ is the two-dimensional density, $J^i$ is the two-dimensional flow, and $\nabla_i^{(2)}$ 
is the two-dimensional covariant derivative.
The anomalous term is obtained by the expansion of $\epsilon$.
We find that this term has a Gaussian and mean curvature dependence and can be written 
as the total divergence of some geometric flow $J^i_{G}$. In total, we have

$$
\frac{\partial \rho}{\partial t} + \nabla_i^{(2)} (J^i+J_G^i)=0.
$$
This fact holds in both classical and quantum mechanics 
when we confine particles to a curved surface with a small thickness.
\end{abstract}
\pacs{87.10.-e, 02.40.Hw, 02.40.Ma, 82.40.Ck}
\maketitle

\section{Geometrical Tools}

The particle motion on a given curved surface $M^2$ is an interesting problem 
in a wide range fields in physics, for example, 
the diffusion or brownian motion \cite{diffusion_equation}, \cite{ogawa_thickness},
the fluid dynamics \cite{flow}, the pattern formation \cite{pattern_formation}, 
Josephson effect \cite{Josephson}, morphogenesis of melanoma in medical science 
\cite{medical_science}, chemical biology \cite{chemical_biology}, quantum mechanics 
\cite{da Costa}, \cite{ogawa_fujii}, \cite{ogawa}, \cite{fujii}, and so on. 
In our study we first explain the diffusion process, 
and then study the quantum mechanics on such a manifold.

  Usually the classical dynamics of particles on such a manifold is expressed 
just by changing the Laplacian to the Laplace-Beltrami operator in the diffusion equation; 
however, when the surface has a thickness $\epsilon$, 
i.e., the configuration space is $M^2 \times R^1$, (see FIG.1.) 
the situation is not simple \cite{ogawa_thickness}.

To make the problem concrete, we first introduce a two-dimensional curved manifold $\Sigma$ in $R^3$, 
and we also introduce two similar copies of $\Sigma$ denoted $\Sigma'$ and $\tilde{\Sigma}$ 
and place them on both sides of $\Sigma$ at a small distance of $\epsilon/2$.

\begin{figure}[h]
\begin{center}
\includegraphics[width=4.5cm]{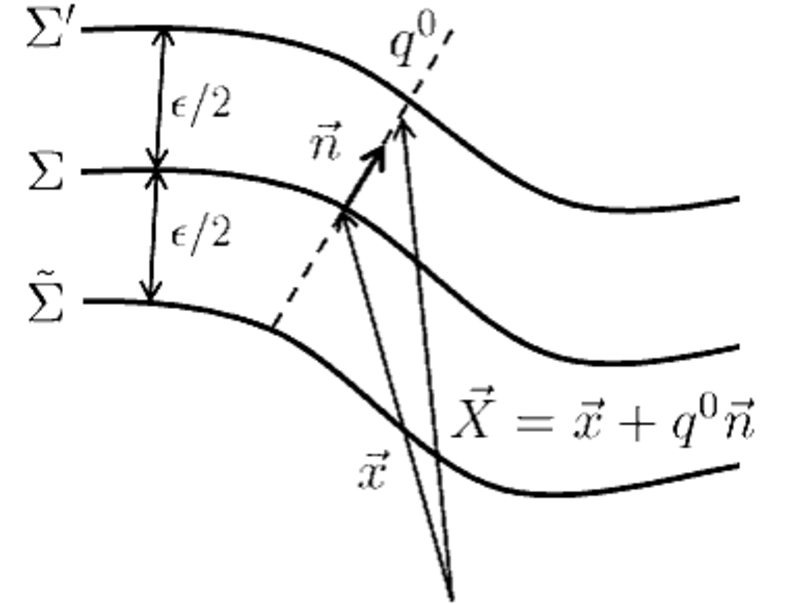}
\end{center}
\caption{curved surface with thickness $\epsilon$}
\end{figure}

Our physical space is between these two surfaces $\Sigma'$ and $\tilde{\Sigma}$.
We set the coordinate system as follows.

$\vec{X}$ is a Cartesian coordinate in $R_3$.
$\vec{x}$ is a Cartesian coordinate that specifies the points on $\Sigma$.
$q^i$ is a curved coordinate on $\Sigma$, where the small Latin indices $i,j,k,\cdots$ run from 1 to 2.
$q^0$ is the coordinate in $R_3$ normal to $\Sigma$.
Furthermore by using the normal unit vector $\vec{n}(q^1,q^2)$ on $\Sigma$ at the point $(q^1,q^2)$,
we can identify any point between the two surfaces  $\Sigma'$ and $\tilde{\Sigma}$ 
by the following thin-layer approximation:

\begin{equation}
\vec{X}(q^0,q^1,q^2) = \vec{x}(q^1,q^2) + q^0 \vec{n}(q^1,q^2),
\end{equation}
where $ -\epsilon/2 \leq q^0 \leq \epsilon/2 $.

Then we obtain the curvilinear coordinate system between two surfaces ($\subset R_3$) by using the coordinate $q^{\mu}=(q^0,q^1,q^2)$ and the metric $G_{\mu\nu}$. 
(Hereafter, Greek indices $\mu, \nu,\cdots$ run from 0 to 2.)

\begin{equation}
G_{\mu\nu}\equiv \frac{\partial \vec{X}}{\partial q^{\mu}} \cdot \frac{\partial \vec{X}}{\partial q^{\nu}}.
\end{equation}

Each part of $G_{\mu\nu}$ is expressed as follows:

\begin{equation}
G_{ij}=g_{ij} + q^0 (\frac{\partial \vec{x}}{\partial q^{i}} \cdot \frac{\partial \vec{n}}{\partial q^{j}}
+ \frac{\partial \vec{x}}{\partial q^{j}} \cdot \frac{\partial \vec{n}}{\partial q^{i}}) + (q^0)^2 \frac{\partial \vec{n}}{\partial q^{i}} \cdot \frac{\partial \vec{n}}{\partial q^{j}},
\end{equation}
where
\begin{equation}
g_{ij} \equiv \frac{\partial \vec{x}}{\partial q^{i}} \cdot \frac{\partial \vec{x}}{\partial q^{j}}
\end{equation}
is the metric (first fundamental tensor) on $\Sigma$. 
Hereafter, the indices $i,j,k \cdots$ are lowered or raised by $g_{ij}$ and its inverse $g^{ij}$.
We also obtain

\begin{equation}
G_{0i}=G_{i0}=0, ~~G_{00}=1.
\end{equation}

We can perform the calculation by using new variables.
We first define the tangential vector to $\Sigma$ by

\begin{equation}
\vec{B}_k = \frac{\partial \vec{x}}{\partial q^{k}}.
\end{equation}

Note that $\vec{n} \cdot \vec{B}_k =0$.
Then we obtain two relations:

the Gauss equation
\begin{equation}
\frac{\partial \vec{B}_i}{\partial q^j} = - \kappa_{ij} \vec{n} + \Gamma^k_{ij} \vec{B}_k
\end{equation}

and the Weingarten equation
\begin{equation}
\frac{\partial \vec{n}}{\partial q^j} = \kappa_j^m \vec{B}_m,
\end{equation}

where 
$$\Gamma^k_{ij} \equiv \frac{1}{2} g^{km}(\partial_i g_{mj} + \partial_j g_{im} - \partial_m g_{ij}).$$

$\kappa_{ij}$ is a symmetric tensor called the Euler-Schauten tensor, or the second fundamental tensor defined by the above two equations.
Furthermore, the mean curvature is given by

\begin{equation}
\kappa = g^{ij} \kappa_{ij},
\end{equation}

and the Ricci scalar $R$ (Gaussian curvature) is defined by

\begin{equation}
R/2 \equiv \det(g^{ik} \kappa_{kj})= \det(\kappa^i_j) = \frac{1}{2}(\kappa^2 -\kappa_{ij} \kappa^{ij}). \label{eq:riemann}
\end{equation}

Then we have the following formula for the metric of curvilinear coordinates in a neighborhood of $\Sigma$:
\begin{equation}
G_{ij}=g_{ij} + 2 q^0 \kappa_{ij} + (q^0)^2 \kappa_{im} \kappa^m_j. \label{eq:total_metric}
\end{equation}

Now we have a total metric tensor such as
\begin{equation}
G_{\mu\nu} =
\left(
\begin{array}{cc}
1 & ~~~0~~~\\
0 &G_{ij}
\end{array}
\right).
\end{equation}

By using the above relations, 
we can construct the diffusion equation and Schr\"{o}dinger equation in our environment.

\section{Classical Diffusion field}

For the classical diffusion field, the problem has already been solved and discussed 
in \cite{ogawa_thickness}.
However, to compare our results with those in quantum mechanics, 
which will be discussed in the next section, 
we briefly sketch the essential part of the results here at the expense of repetition.

We denote the three-dimensional diffusion field in our space 
between $\Sigma'$ and $\tilde{\Sigma}$ as $\phi^{(3)}$, 
which satisfies the usual three-dimensional diffusion equation and normalization condition.
\begin{eqnarray}
&& \frac{\partial \phi^{(3)}}{\partial t} = D \Delta^{(3)} \phi^{(3)} \label{eq:diff}\\
n &=& \int \phi^{(3)}(q^0,q^1,q^2)  \sqrt{G}~ d^3 q
\end{eqnarray}

where, $D$ is the diffusion constant, $\Delta^{(3)}$ is the three dimensional Laplace Beltrami operator, 
$n$ is the number of particles, and $G = \det(G_{\mu\nu}) = \det (G_{ij})$.
When $\epsilon \to 0$, the theory reduces to the two-dimensional theory.
Our aim is to construct an effective two-dimensional equation 
from three-dimensional equation with a small but finite $\epsilon$.
The effective two-dimensional diffusion field $\phi^{(2)}(q^1,q^2)$ 
should satisfy a normalization condition such as
\begin{equation}
n = \int \phi^{(2)}(q^1,q^2)  \sqrt{g}~ d^2 q ,
\end{equation}
where  $g = \det(g_{ij})$.

From the two normalization conditions, we obtain
\begin{eqnarray}
n &=& \int \phi^{(3)}(q^0,q^1,q^2)  \sqrt{G}~ d^3 q,\nonumber \\
&=& \int [\int_{-\epsilon/2}^{\epsilon/2} d q^0  (\phi^{(3)} \sqrt{G/g})] ~ \sqrt{g} ~ d^2 q, \nonumber\\
&=& \int \phi^{(2)}(q^1,q^2)  \sqrt{g}~ d^2 q.\nonumber
\end{eqnarray}
Therefore, we obtain the relationship
\begin{equation}
\phi^{(2)}(q^1, q^2) = \int_{-\epsilon/2}^{\epsilon/2} \tilde{\phi}^{(3)} d q^0, \label{eq:relation}
\end{equation}
where
\begin{equation}
\tilde{\phi}^{(3)} \equiv \phi^{(3)} \sqrt{G/g}. \label{eq:definition}
\end{equation}

We further suppose the local equilibrium condition that there is no diffusion flow 
in the normal direction to the layer \cite{ogawa_thickness}.
\begin{equation}
0 = \frac{\partial \phi^{(3)}}{\partial q^0} \label{eq:suppose}
\end{equation}

Then from (\ref{eq:relation}), (\ref{eq:definition}), and (\ref{eq:suppose}),
we can write $\tilde{\phi}^{(3)}$ in terms of $\phi^{(2)}$.
\begin{eqnarray}
\tilde{\phi}^{(3)} &=& \frac{1}{N}  (G/g)^{1/2} \phi^{(2)}(q^1,q^2),\label{eq:relation2} \\
N &\equiv & \int_{-\epsilon/2}^{\epsilon/2}  (G/g)^{1/2} dq^0. 
\end{eqnarray}

The physical interpretation of (\ref{eq:suppose}) is as follows.
The diffusion in the normal direction to the surface may reach equilibrium in a short time of 
 $ \delta t \sim \epsilon^2/D $. 
Therefore, if we consider the diffusion system with the time scale $t >> \delta t$, 
we can assume equilibrium in this direction, expressed by (\ref{eq:suppose}), all the time. In other words, the diffusion on the surface occurs always while satisfying the equilibrium condition in the normal direction in this time scale. 
 
We multiply the diffusion equation (\ref{eq:diff}) by $\sqrt{G/g}$ , 
and integrate with respect to $q^0$, and by using  (\ref{eq:relation}) and  (\ref{eq:relation2}), 
we obtain the final form of the equation up to 
$ {\cal O}(\epsilon^2)$ as

\begin{flushleft}
\begin{eqnarray}
\frac{\partial \phi^{(2)}}{\partial t} 
&=& D \Delta^{(2)} \phi^{(2)}  + \tilde{D} g^{-1/2} \frac{\partial}{\partial q^i} ~ g^{1/2} \nonumber\\
&\times& \{ (3 \kappa^{im} \kappa_m^j -2 \kappa \kappa^{ij}) \frac{\partial}{\partial q^j} 
- \frac{1}{2} g^{ij} \frac{\partial R}{\partial q^j} \} \phi^{(2)},~~~
\end{eqnarray}
\end{flushleft}
where $\tilde{D} = \frac{\epsilon^2}{12} D$.\\

We can rewrite the diffusion equation in the form
\begin{eqnarray}
-\frac{\partial \phi^{(2)}}{\partial t} &=& \nabla_i^{(2)} (J^i + J_G^i),\nonumber\\
&=& g^{-1/2} \frac{\partial }{\partial q^i} ~g^{1/2} (J^i + J_G^i),
\end{eqnarray}
where $\nabla_i^{(2)}$ is the two-dimensional covariant derivative,
the normal diffusion flow is
\begin{equation}
J^i = -D g^{ij} \frac{\partial \phi^{(2)}}{\partial q^j},
\end{equation}
and the anomalous diffusion flow is
\begin{equation}
J_G^i = -\tilde{D} ~[ ~(3 \kappa^{im} \kappa_m^j - 2\kappa \kappa^{ij})
\frac{\partial \phi^{(2)}}{\partial q^j} 
- \frac{1}{2} g^{ij} \frac{\partial R }{\partial q^j}\phi^{(2)}]. \label{eq:anom}
\end{equation}

\section{Quantum mechanics in the same geometry}
The quantum mechanics on curved manifold embedded 
in higher dimensional Euclidean space has a long history.
The reason is the following. When we consider the curved space from the outset, 
we can not construct the quantum theory without ambiguity.
This is coming from the operator ordering problem existing in quantization rule.
One method to avoid the problem is to consider the extrinsic world: i.e. 
curved manifold is embedded in higher dimensional Euclidean space.
The quantization is done in Euclidean space, and 
then we confine the particle onto the submanifold.
There are two methods for such quantization with constraint.
One is by using the Dirac's method \cite{ogawa_fujii}, 
and the another method is so called confining potential method \cite{da Costa}.
The Dirac method is the quantization rule for such a constrained dynamical system,
and there is no degree of freedom into normal ($q^0$) direction. 
On the other hand, In the confining potential method, quantization is done 
in external euclidean space and constraint is given by potential function for example:
\begin{equation}
V(q^0) = \left\{
         \begin{array}{ll}
              0 & (-\epsilon/2 < q^0 < +\epsilon/2) \\
              \infty & (q^0 = \pm \epsilon/2)
         \end{array}
         \right.
\end{equation}
Then we take the limit $\epsilon \to 0$.
In both approaches, we obtain the geometrical potential with $\hbar^2$ ,
but are different. The reason of its different potential is discussed by Ogawa 
\cite{ogawa}. Furthermore, we have an additional geometrical gauge field in the case of 
confining potential method, for example, $O(2)$ gauge field appears for the curved line 
in 3 dimensional Euclidean space, 
and in general case: d-dimensional curved submanifolds embedded 
in higher n-dimensional Euclidian space ($n-d \geq 2$), 
we have $O(n-d)$ gauge field \cite{fujii}.
In this manuscript, we work with this confining potential method,
 but without taking the limit $\epsilon \to 0$. 
Between the two surfaces $\Sigma'$ and $\tilde{\Sigma}$,
 our basic equation is the Schr\"{o}dinger equation 
which is written by using curvilinear coordinates in a three-dimensional space.

\begin{equation}
i\hbar \frac{\partial}{\partial t} \psi 
= [-\frac{\hbar^2}{2m} \Delta^{(3)} + V(q) ] \psi,  \label{eq:sch1}
\end{equation}
where the form of the Laplace-Beltrami operator is 
$$ \Delta^{(3)} \equiv G^{-1/2} \partial_{\mu} G^{1/2} G^{\mu \nu} \partial_\nu,$$
and we suppose that $V$ depends on neither $t$ nor $q^0$.

 Starting from this wave function $\psi$, 
we construct the effective two-dimensional theory \cite{da Costa} - \cite{fujii}.
 From the normalization condition we obtain

\begin{equation}
1 = \int |\psi |^2 \sqrt{G} d^3q = \int [ \int_{-\epsilon/2}^{+\epsilon/2} |\psi|^2 \sqrt{\frac{G}{g}} ~~dq^0] \sqrt{g} d^2q.
\end{equation}

Our effective two-dimensional wave function $\phi$ should satisfy
\begin{eqnarray}
|\phi(q^1, q^2)|^2 &=&  \int_{-\epsilon/2}^{+\epsilon/2} |\psi|^2 \sqrt{\frac{G}{g}} ~~dq^0,\\
1 &=& \int |\phi|^2 \sqrt{g}~ d^2q.
\end{eqnarray}

Then how can we obtain the Schr\"{o}dinger equation for $\phi$ ? 
To solve this problem, we first define a new variable $\tilde{\psi}$ as
\begin{equation}
\tilde{\psi} \equiv (G/g)^{1/4} \psi  \label{eq:tilde}
\end{equation}
with
\begin{equation}
|\phi(q^1, q^2)|^2 =  \int_{-\epsilon/2}^{+\epsilon/2} |\tilde{\psi}|^2  ~~dq^0,\\
\end{equation}

Furthermore we suppose that it is possible to separate the variables.

\begin{eqnarray}
\tilde{\psi} &=& \phi(q^1, q^2, t)~ \chi(q^0,t), \label{eq:sov}\\
1 &=& \int_{-\epsilon/2}^{+\epsilon/2} |\chi|^2  ~dq^0.
\end{eqnarray}

Then we can construct the equation for $\phi$ as follows.

First, we construct the Schr\"{o}dinger equation for $\tilde{\psi}$. This has the same form as (\ref{eq:sch1}) except that the Laplace-Beltrami operator is changed to the following operator.

\begin{equation}
\tilde{\Delta}^{(3)} \equiv (G/g)^{1/4} \Delta^{(3)} (G/g)^{-1/4}.
\end{equation}

Using one of the tools in Appendix A, this operator can be expanded as

\begin{eqnarray}
\tilde{\Delta}^{(3)} &=& \Delta^{(2)} + \frac{\partial^2}{\partial (q^0)^2} + V_0 
+ q^0 V_1 + (q^0)^2 V_2 \nonumber\\
&&  + q^0 \hat{A}_1 + (q^0)^2 \hat{A}_2 + {\cal O}((q^0)^3), \label{eq:laplace}
\end{eqnarray}
where $V_0, V_1, V_2,\hat{A}_1,$ and  $\hat{A}_2$ are given by

\begin{eqnarray}
V_0 &=& \frac{1}{4} (\kappa^2 - 2R),\\
V_1 &=& \kappa ( R-\frac{\kappa^2}{2}) - \frac{1}{2} \Delta^{(2)} \kappa,\\
V_2 &=& \frac{3}{4} \kappa^4 - \frac{7}{4}\kappa^2 R + \frac{1}{2} R^2 + \frac{1}{2} \kappa \Delta^{(2)} \kappa \nonumber\\
&& + \frac{1}{4} g^{ij} (\partial_i \kappa)(\partial_j \kappa) + \nabla_i (\kappa^{ij} \partial_j \kappa)\nonumber\\
&& - \frac{1}{4} \Delta^{(2)} R,\\
\hat{A}_1 &=& -2 \nabla_i \kappa^{ij} \partial_j,\\
\hat{A}_2 &=& 3 \nabla_i \kappa^{ik}\kappa_k^j  \partial_j.
\end{eqnarray}

Then our equation for $\tilde{\psi}$ is given as follows:

\begin{eqnarray}
i\hbar \frac{\partial \tilde{\psi}}{\partial t} &=& -\frac{\hbar^2}{2m}  [ \Delta^{(2)} + \frac{\partial^2}{\partial (q^0)^2} + V_0 + q^0 V_1 + (q^0)^2 V_2 \nonumber\\
&& +  q^0 \hat{A}_1 + (q^0)^2 \hat{A}_2]  \tilde{\psi} + V \tilde{\psi}, \label{eq:sch2}
\end{eqnarray}
where we have omitted ${\cal O}((q^0)^3)$ terms for small $q^0$.

We treat this system by the perturbation method.
The Hamiltonian can be written as

\begin{eqnarray}
\hat{H} &=& \hat{H}_0 + \hat{H}_I,\\
\hat{H}_0 &=& -\frac{\hbar^2}{2m}  [ \Delta^{(2)} + \frac{\partial^2}{\partial (q^0)^2} + V_0] + V,\\
\hat{H}_I &=& -\frac{\hbar^2}{2m}  [ q^0 (V_1 + \hat{A}_1) + (q^0)^2 (V_2 +\hat{A}_2)].
\end{eqnarray}

As an eigenfunction of $\hat{H}_0$, we introduce $\tilde{\chi}_N$ and $\tilde{\phi}_k$, where
\begin{eqnarray}
 &&-\frac{\hbar^2}{2m} \frac{\partial^2}{\partial (q^0)^2} \tilde{\chi}_N = E_N ~ \tilde{\chi}_N ,\\
 &&[ -\frac{\hbar^2}{2m}  ( \Delta^{(2)} + V_0) + V] \tilde{\phi}_k = \lambda_k ~ \tilde{\phi}_k . \label{eq:sch3}
\end{eqnarray}

The eigenfunction $\tilde{\chi}_N$ should also satisfy the Dirichlet boundary condition
$\tilde{\chi}_N (q^0 = \pm \epsilon/2)=0$. 
Then the time dependent orthonormal eigenfunction $\chi_N ~(N=1,2,3, \cdots)$ is given as
\begin{equation}
\chi_N = \tilde{\chi}_N ~ e^{-i E_N t/\hbar},
\end{equation}
where
\begin{eqnarray}
\tilde{\chi}_N 
&=& \left\{
\begin{array}{l}
\sqrt{\frac{2}{\epsilon}} \cos (N \pi q^0/\epsilon) ~~~N=\mbox{odd},\\
\sqrt{\frac{2}{\epsilon}} \sin (N \pi q^0/\epsilon) ~~~N=\mbox{even}. 
\end{array}\right. \label{eq:chi}
\end{eqnarray}

and
$$E_N = \frac{\hbar^2 \pi^2}{2m\epsilon^2} N^2.$$

Note that
\begin{equation}
\int_{-\epsilon/2}^{+\epsilon/2} \tilde{\chi}^*_M \tilde{\chi}_N ~dq^0  
\equiv ~<M|N> = \delta_{MN}.
\end{equation}
\

We also assume the existence of a time-dependent orthonormal eigenfunction $\phi_k$ and eigenvalue $\lambda_k$ as a solution of the equation (\ref{eq:sch3}) without giving their explicit form as follows:
\begin{equation}
\phi_k = \tilde{\phi}_k ~ e^{-i \lambda_k t/\hbar},
\end{equation}
and
\begin{equation}
\int \tilde{\phi}^*_k  ~\tilde{\phi}_j \sqrt{g} ~dq^2  \equiv ~(k|j) = \delta_{kj}.
\end{equation}

Therefore, the eigenfunction of $\hat{H}_0$ is given by the direct product of $\chi$ and $\phi$,
\begin{equation}
\hat{H}_0 \left| M \right\rangle \left| k \right ) = (E_M + \lambda_k) \left| M \right\rangle \left| k \right ).
\end{equation}

Let us we consider the perturbation theory for $\hat{H}_I$.
The important point is that a transition of the quantum number $N,M,\cdots$ does not occur 
in our perturbation up to ${\cal O}(\epsilon^2)$. 
Because if we consider the transition like $N \to M$, 
we have a correction to the state $|N>$ by $|M>$ with coefficient such as
\begin{equation}
\frac{\left\langle M \right | \hat{H}_I \left| N \right\rangle}{E_N-E_M} \sim 
\frac{\cal O(\epsilon)}{1/\epsilon^2} = {\cal O}(\epsilon^3).
\end{equation}
(The precise discussions are given in Appendix B.)
In this approximation, our effective Hamiltonian can be expressed as
\begin{eqnarray}
\hat{H}_N &\equiv& \left\langle N \right |\hat{H} \left| N \right\rangle \nonumber\\
&=& E_N -\frac{\hbar^2}{2m}  [ \Delta^{(2)}  + V_0] + V  \nonumber \\
&-&\frac{\hbar^2}{2m}  [\left\langle N \right | q^0 \left| N \right\rangle(V_1 + \hat{A}_1)\nonumber\\
&& ~~~ + \left\langle N \right |(q^0)^2 \left| N \right\rangle(V_2 +\hat{A}_2)].
\end{eqnarray}

The expectation values of $q^0$ and $(q^0)^2$ are
\begin{eqnarray}
\left\langle N \right | q^0 \left| N \right\rangle &=&  0, \\
\left\langle N \right |(q^0)^2 \left| N \right\rangle &=&  \frac{\epsilon^2}{12}(1 - \frac{6}{\pi^2 N^2})  \equiv \tilde{\epsilon}^2.            
\end{eqnarray}

Then we employ the separation of variables method to obtain
\begin{equation}
\tilde{\psi} = \chi_N(q^0,t)~ \phi(q^1, q^2, t). \label{eq:form}
\end{equation}

By substituting (\ref{eq:form}) into (\ref{eq:sch2}), multiplying by $\chi^*_N$ and integrating with respect to $q^0$, we obtain
\begin{equation}
E_N \phi + i\hbar \dot{\phi}= \hat{H}_N \phi.
\end{equation}

This leads to the following approximate Schr\"{o}dinger equation for $\phi$.
\begin{eqnarray}
 i \hbar \frac{\partial \phi}{\partial t} &=& [-\frac{\hbar^2}{2m}  \{ \Delta^{(2)}  + V_0 \nonumber\\
&& ~~~~~+ \tilde{\epsilon}^2 (V_2 +\hat{A}_2)  \} + V ] \phi. \label{eq:Sch}
\end{eqnarray}

From this Schr\"{o}dinger equation, we obtain
\begin{eqnarray}
& -& \frac{\partial}{\partial t} |\phi|^2 = \frac{\hbar}{2mi} \{ (\phi^* \Delta^{(2)} \phi - \phi \Delta^{(2)} \phi^*) \nonumber\\
&& ~~~~~~~~~~~~~~~+  \tilde{\epsilon}^2 (\phi^* \hat{A}_2 \phi -  \phi \hat{A}_2 \phi^*)\}\nonumber\\
&=&  \frac{\hbar}{2mi}\nabla_i \{ (g^{ij} + 3 \tilde{\epsilon}^2 \kappa^{ik} \kappa^j_k) (\phi^* \partial_j \phi - \phi \partial_j \phi^*) \} \nonumber\\
&=& \nabla_i (J^i + J_G^i),
\end{eqnarray}
where
\begin{eqnarray}
J^i &\equiv &  \frac{\hbar}{2mi} \{g^{ij} (\phi^* \partial_j \phi - \phi \partial_j \phi^*) \}, 
\label{eq:current}\\
J_G^i &\equiv &  \frac{3 \hbar \tilde{\epsilon}^2 }{2mi} ~ \kappa^{ik} \kappa^j_k~(\phi^* \partial_j \phi - \phi \partial_j \phi^*) .\label{eq:anomalous_current}
\end{eqnarray}
where, $\tilde{\epsilon}$ has $N$ dependence and
$$\tilde{\epsilon}^2 =\frac{\epsilon^2}{12}(1 - \frac{6}{\pi^2 N^2}). $$
For low energy physics, we take $N=1$. The explicit example is shown in appendix C.

\section{Conclusion}
We have discussed the conservation law in an effective two-dimensional system 
between two curved surfaces $\Sigma'$ and $\tilde{\Sigma}$ separated by a small distance 
$\epsilon$. 
We found that the anomalous flow depends on the curvature of the surface $\Sigma$. 
In the classical diffusion process, we have

\begin{equation}
J_G^i = -\frac{\epsilon^2 D}{12} ~[ ~(3 \kappa^{im} \kappa_m^j - 2\kappa \kappa^{ij})
\frac{\partial \phi^{(2)}}{\partial q^j} 
- \frac{1}{2} g^{ij} \frac{\partial R }{\partial q^j}\phi^{(2)}]
\end{equation}
 as shown in \cite{ogawa_thickness}.
 
In the quantum process we instead obtained

\begin{equation}
J_G^i =  \frac{\hbar \epsilon^2 }{8mi} (1-\frac{6}{\pi^2 N^2})~ \kappa^{ik} \kappa^j_k~(\phi^* \partial_j \phi - \phi \partial_j \phi^*),
\end{equation}
where $N$ is the quantum number which defines the energy state of motion in the $q^0$ direction.
The classical anomalous flow and quantum mechanical anomalous flow are somewhat similar. 
Both start from ${\cal O}(\epsilon^2 \kappa^2)$ and are proportional to the gradient of the field 
except the last term in the classical flow.
The recent nano-technology made it possible to fabricate complicate devices. 
Then this kind of anomalous flow might play an important role 
in such a ``geometrical" device.

\section{Appendix}

\subsection{Geometrical Tools}

From the definitions of the metric (\ref{eq:total_metric}) and the Ricci scalar
 (\ref{eq:riemann}), we can construct the following geometrical quantities:

\begin{equation}
G \equiv \det{G_{ij}} = g + 2g \kappa q^0 + g(\kappa^2 +R) (q^0)^2 + {\cal O}((q^0)^3).
\end{equation}
The inverse metric of $G_{ij}$ is given as
\begin{equation}
G^{ij} = g^{ij} -2\kappa^{ij} q^0 + \frac{3}{2} (2\kappa \kappa^{ij} -R g^{ij}) (q^0)^2 +  {\cal O}((q^0)^3).
\end{equation}

Furthermore,
\begin{equation}
(\kappa^{-1})^{ij} = \frac{2}{R} (\kappa g^{ij} - \kappa^{ij}),
\end{equation}
and from this relationship we obtain

\begin{eqnarray}
\frac{1}{2}R g^{ij} &=& \kappa \kappa^{ij} - \kappa^i_m \kappa^{mj},\\
R &=& \kappa^2 - \kappa^i_j \kappa^j_i =2 \det(\kappa^i_j).
\end{eqnarray}

These relationships are used to decompose the Laplace-Beltrami operator (\ref{eq:laplace}).
\\

\subsection{Perturbation Theory}

As is well known from the texts on quantum mechanics, 
by utilizing energy eigenvalue of $\hat{H}_0$, and the state vector $\left| n \right\rangle$, the corrected energy eigenvalue and state vector in the first-order perturbation
 are generally given as
\begin{eqnarray}
E'_n &=& E_n + \left\langle n \right|\hat{H}_I\left| n \right\rangle ,\\
\left| n \right\rangle' &=& \left| n \right\rangle + \sum_{k \neq n} \frac{\left\langle k \right| \hat{H}_I\left| n \right\rangle}{E_n-E_k} \left| k \right\rangle.
\end{eqnarray}

In our case, we have two degrees of freedom $\chi$ and $\phi$. Their natural extension are given as

\begin{eqnarray}
&& E'_{N,k} = E_N + \lambda_k + \left\langle N \right |  (k| \hat{H}_I |k)  \left| N \right\rangle, \label{eq:pert1} \\
&& \left| N,k \right\rangle'= \left| N \right\rangle \left| k \right ) \nonumber\\
&&+ \sum_{(M,j) \neq (N,k)}\frac{\left\langle M \right | (j| \hat{H}_I |k) \left| N \right\rangle}{(E_N+\lambda_k)-(E_M + \lambda_j)}  \left| M \right\rangle \left| j \right ).
\end{eqnarray}

When $M \neq N$, the denominator of the perturbation energy is the order of 
${\cal O}(\epsilon^{-2})$ , so that the term 
\begin{equation}
\frac{\left\langle M \right | \hat{H}_I \left| N \right\rangle}{E_N-E_M} \sim 
\frac{\cal O(\epsilon)}{1/\epsilon^2} = {\cal O}(\epsilon^3).
\end{equation}

This means that we can ignore the change of the quantum number $N$ under the perturbation up to 
${\cal O}(\epsilon^2)$. Therefore, we only need to consider the case of $M=N$:

\begin{eqnarray}
\left| N,k \right\rangle'& \cong & \left| N \right\rangle \left| k \right ) \nonumber\\
&+& \sum_{j \neq k}\frac{\left\langle N \right | (j| \hat{H}_I |k) \left| N \right\rangle}{\lambda_k- \lambda_j}  \left| N \right\rangle \left| j \right ). \label{eq:pert2}
\end{eqnarray}

Our perturbation theory is given by (\ref{eq:pert1}) and (\ref{eq:pert2}).
The Hamiltonian that directly leads to (\ref{eq:pert1}) and (\ref{eq:pert2}) is

\begin{eqnarray}
\hat{H}_N &\equiv& \left\langle N \right |\hat{H} \left| N \right\rangle \nonumber\\
&=& E_N -\frac{\hbar^2}{2m}  [ \Delta^{(2)}  + V_0] + V  \nonumber \\
&-&\frac{\hbar^2}{2m}  [\left\langle N \right | q^0 \left| N \right\rangle(V_1 + \hat{A}_1)\nonumber\\
&& ~~~ + \left\langle N \right |(q^0)^2 \left| N \right\rangle(V_2 +\hat{A}_2)].
\end{eqnarray}

\subsection{An Example}
To consider the physical meaning of geometric flow, let us show one simple example: 
bending ribbon with thickness $\epsilon$ as seen in figure 2.

\begin{figure}[h]
\begin{center}
\includegraphics[width=4cm]{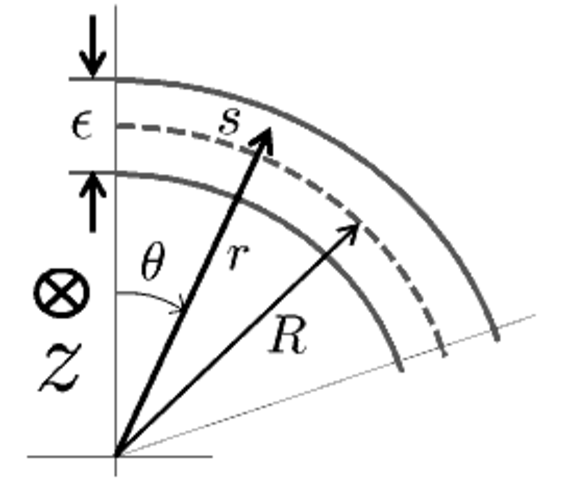}
\end{center}
\caption{Bending ribbon (side view) with thickness $\epsilon$.}
\end{figure}

The physical space is inner ribbon with $0<z<L$, and $R-\epsilon/2 < r < R+\epsilon/2$. 
Our starting equation is an usual three dimensional Schr\"odinger equation 
(\ref{eq:sch1}). Then we obtain the conservation law,
\begin{eqnarray}
&& \nabla_\mu \tilde{J}^\mu = - \frac{\partial \tilde{\rho}}{\partial t},\\
&& \tilde{J}^\mu = \frac{\hbar}{2mi} G^{\mu\nu} ( \psi^* \nabla_\nu \psi - \psi \nabla_\nu \psi^*),
\end{eqnarray}
where $\tilde{J}$ and $\tilde{\rho}$ are three dimensional flow and density respectively.
We utilize the cylindrical coordinates $(r, \theta, z)$ and
\begin{eqnarray}
ds^2_{(3)} &=& dr^2 + r^2 d\theta^2 +dz^2,\\
G_{\mu\nu} &=& \mbox{diag}(1, r^2, 1), \sqrt{G} = r.
\end{eqnarray}

\begin{eqnarray}
\tilde{J}^r &=& \frac{\hbar}{2mi} (\psi^* \partial_r \psi - \psi \partial_r \psi^*),\\
\tilde{J}^\theta &=& \frac{\hbar}{2m r^2 i} (\psi^* \partial_\theta \psi - \psi \partial_\theta \psi^*),\\
\tilde{J}^z &=& \frac{\hbar}{2mi} (\psi^* \partial_z \psi - \psi \partial_z \psi^*).
\end{eqnarray}
Then the conservation law gives
\begin{equation}
\partial_\theta \tilde{J}^\theta + \partial_r \tilde{J}^r +\partial_z \tilde{J}^z 
+ \frac{1}{r} \tilde{J}^r =-\partial_t \tilde{\rho}. \label{eq:cons}
\end{equation}

The volume element is given by
\begin{equation}
dV = r dr d\theta dz = (r/R) dr ds dz,
\end{equation}
where $ds=R d\theta$. We then integrate both hand sides of (\ref{eq:cons}) by $(r/R)dr$ in a region $R-\epsilon/2 \sim R+\epsilon/2$ and then we obtain two dimensional conservation law.
\begin{equation}
\partial_s J^s_{tot}  +\partial_z J^z_{tot} = -\partial_t \rho,~~~ \int \rho ~ds dz =1,
\end{equation}
where
\begin{eqnarray}
J^s_{tot} \equiv \int_{R-\epsilon/2}^{R+\epsilon/2}r \tilde{J}^\theta ~dr,\\
J^z_{tot} \equiv \int_{R-\epsilon/2}^{R+\epsilon/2} r \tilde{J}^z/R ~dr,\\
\rho \equiv \int_{R-\epsilon/2}^{R+\epsilon/2} r \tilde{\rho}/R ~dr,
\end{eqnarray}
where the boundary condition $\tilde{J}^r =0$ at $r= R \pm \epsilon/2$ and $ds=R d\theta$ are utilized.
We have
\begin{equation}
J^s_{tot} = \frac{\hbar}{2mi}\int_{R-\epsilon/2}^{R+\epsilon/2} \frac{dr}{r} (\psi^* \partial_\theta \psi - \psi \partial_\theta \psi^*).
\end{equation}

The separation of variable (\ref{eq:tilde}), (\ref{eq:sov}) gives
\begin{equation}
\psi (r, \theta, z) = (g/G)^{1/4} \varphi(\theta, z) \chi(r) = \sqrt{\frac{R}{r}}\varphi(\theta, z) \chi(r) \label{eq:sv},
\end{equation}
where
$$ \int dr ~|\chi|^2 =1, ~~ \int R d\theta dz ~|\varphi|^2=\int ds dz~|\varphi|^2=1,$$
$$ ds^2_{(2)} = R^2 d\theta^2 + dz^2, ~~ g_{ij} = \mbox{diag} (R^2,1),~~ \sqrt{g} = R.$$

Then we obtain

\begin{equation}
J^s_{tot} = \frac{\hbar R^2}{2mi}\int_{R-\epsilon/2}^{R+\epsilon/2} \frac{dr}{r^2}~ |\chi(r)|^2 (\varphi^* \partial_s \varphi - \varphi \partial_s \varphi^*).
\label{eq:toy_geometric_flow1}
\end{equation}

Equation (\ref{eq:chi}) gives the explicit form of $\chi$. 
By using $$x \equiv q^0/R =(r-R)/R,$$
we obtain
\begin{eqnarray}
J^s_{tot} &=& \frac{\hbar R}{m i\epsilon}\int_{-\epsilon/2R}^{+\epsilon/2R} \frac{dx}{(1+x)^2} \cos^2((N \pi R/\epsilon)x), \nonumber \\
&& ~~~~~~~~~~~~~~ \times (\varphi^* \partial_s \varphi - \varphi \partial_s \varphi^*),
\end{eqnarray}
for odd $N$. For even $N$ we just change $\cos$ to $\sin$.

We expand the integrand as
\begin{equation}
\frac{1}{(1+x)^2} = 1 -2x + 3x^2 + \cdots.\label{eq:exp}
\end{equation}
Then we obtain
\begin{equation}
J^s_{tot} = \frac{\hbar}{2mi}(1+3\tilde{\epsilon}^2/R^2 + \cdots) (\varphi^* \partial_s \varphi - \varphi \partial_s \varphi^*),
\end{equation}
for both of even and odd $N$.
If we use coordinate $q^{i} = (s,z)$, we have $g_{ij} =\delta_{ij},~~ \kappa_{ss} =1/R,~~ \kappa_{sz}=\kappa_{zz}=0$. Then we find this is completely equal to $J^s +J^s_G$ appeared 
in (\ref{eq:current}) and (\ref{eq:anomalous_current}).

On the other hand,
\begin{eqnarray}
J^z_{tot} &=& \frac{\hbar}{2mi}(\int_{R-\epsilon/2}^{R+\epsilon/2} dr ~|\chi(r)|^2 ) (\varphi^* \partial_z \varphi - \varphi \partial_z \varphi^*) \nonumber\\
&=& \frac{\hbar}{2mi}(\varphi^* \partial_z \varphi - \varphi \partial_z \varphi^*).
\end{eqnarray}
This is equal to $J^z$ in (\ref{eq:current}) , and we have no geometric flow in this straight direction. 
Geometric flow is contained as the part of the integrand in equation (\ref{eq:toy_geometric_flow1}):
the second order of expansion (\ref{eq:exp}). We can transcribe (\ref{eq:toy_geometric_flow1}) into the form
\begin{eqnarray}
J^s_{tot}&=& \frac{\hbar}{2mi}<(\frac{R}{r})^2> (\varphi^* \partial_s \varphi - \varphi \partial_s \varphi^*), \label{eq:QM}\\
<f> &\equiv& \int_{R-\epsilon/2}^{R+\epsilon/2} f(r) ~|\chi(r)|^2 dr.
\end{eqnarray}


\begin{thebibliography}{99}
\bibitem{diffusion_equation}
Faraudo J., 
{\it Diffusion equation on curved surfaces. I. Theory and application to biological membranes}, 
J. Chem. Phys, {\bf 116} (2002) 5831-5841;
Balakrishnan J., 
{\it Spatial Curvature Effects on Molecular Transport by Diffusion}, 
Phys.Rev.E{\bf 61}, (2000) 4648--4651; 
Gov N., 
{\it Diffusion in curved fluid membranes}, 
Phys. Rev. E {\bf 73} (2006) 041918;
Naji A. and Brown F., 
{\it Diffusion on ruffled membrane surfaces}, 
J. Chem. Phys. {\bf 126} (2007) 235103;
Reister E. and Seifert U., 
{\it Lateral diffusion of a protein on a fluctuating membrane}, 
Europhys. Lett. {\bf 71} (2005) 859-865;
Priego R. C., Villarreal P. C., Jimenez S. E., Alcaraz J. M., 
{\it Brownian motion of free particles on curved surfaces}, 
arXive: 1211.5799v2 (2013);
Villarreal P. C., Balbuena A. V., Alcaraz J. M., Priego R. C., Alvarez S. E., 
{\it A Brownian dynamics algorithm for colloids in curved manifolds},
 J. Chem. Phys. {\bf 140} (2014) 214115.

\bibitem{ogawa_thickness}
Ogawa N., 
{\it Curvature Dependent Diffusion Flow on Surface with Thickness}, 
Phys. Rev. {\bf E81} (2010) 061113; 
Ogawa N., 
{\it Diffusion in a Curved Tube}, 
Phys. Lett. {\bf A377} (2013) 2465--2471;
Ogawa N., 
{\it Diffusion Under Geometrical Constraint}, 
J. Geo. Sym. Phys. {\bf 34}, I. Mladenov (Ed),
Bulgarian Academy of Sciences, Sofia 2014, pp~35--49;
Valdes C. V., 
{\it Effective diffusion on Riemannian fiber bundles}, 
J. M. Phys. {\bf 56} 023507 (2015);
Valdes C. V., {\it Effective diffusion in the region between two surfaces},
Phys. Rev. E {\bf 94},022121 (2016).

\bibitem{flow}
Morris R. G., 
{\it Relaxation and curvature-induced molecular flows within multicomponent membranes}, 
Phys. Rev. {\bf E 89} (2014) 062704;
Shi Q., Chen Y., Xie X.,
{\it Interplay of surface geometry and vorticity dynamics in incompressible flows on curved surfaces}, 
App. Math. Mech. {\bf 38}, 1191-1212 (2017).

\bibitem{pattern_formation}
Venkataraman C., Sekimura T., Gaffney E., Maini P., Madzvamuse A., 
{\it Modeling parr-mark pattern formation during the early development of Amago trout}, 
Phys. Rev. {\bf E 84} (2011) 041923; 
Duran A. L., Valencia L. H. J., Malacara J. B. M., Holek I. S., 
{\it The interplay between phenotypic and ontogenetic plasticities The interplay between phenotypic and ontogenetic plasticities can be assessed using reaction diffusion models : The case of Pseudoplatystoma fishes}, J. Biol. Phys. {\bf 43} (2017) 247-264.

\bibitem{Josephson}
Dobrowolski T., {\it Possible curvature effects in the Josephson junction}, 
Eur. Phys. J. B {\bf 86} (2013) 346--354.

\bibitem{medical_science}
Chatelain C., 
{\it Morphogenesis during early melanoma growth}, Doctor thesis, Universit\'e Pierre et 
Marie Curie (Paris 6) and Laboratoire de Physique Statistique de l' \'Ecole Normale Sup\'erieure (2012);
Balois T., Chatelain C., Amar M. B., 
{\it Patterns in melanocytic lesions: impact of the geometry on growth and 
transport inside the epidermis }, J. R. Soc. Interface {\bf 11} (2014) 0339.

\bibitem{chemical_biology}
Assenda S., Mezzenga R. {\it Curvature and bottlenecks control molecular transport in inverse bicontinuous cubic phases}, 
J. Chem. Phys. {\bf 148}, 054902 (2018);
Campbell E. J., Bagchi P., 
{\it A computational model of amoeboid cell motility in the presence of obstacles}, 
Soft Matter {\bf 14}, 5741-5763(2018).

\bibitem{da Costa}
da Costa R.,
{\it Quantum mechanics of a constrained particle}, 
Phys. Rev. {\bf 23} (1981) 1982--1987;
Tolar J., 
{\it On a quantum mechanical d'Alembert principle},
Lecture Notes in Physics {\bf 313}, ed. H. D. Doever, J. D. Henning, T. D. Raev, 
Springer-Verlag, Berlin, Heidelberg 1988, 268--274.

\bibitem{ogawa_fujii}
Ogawa N., Fujii K., Kobushkin K. P.,
{\it Quantum Mechanics in Riemannian Manifold}, 
Prog. Theor. Phys. {\bf 83} (1990) 894--905; 
Ogawa N., Fujii K., Chepilko N. M., Kobushkin K. P., 
{\it Quantum Mechanics in Riemannian Manifold. II},
Prog. Theor. Phys. {\bf 85} (1991) 1189--1201.

\bibitem{ogawa}
Ogawa N.,
{\it The Difference of Effective Hamiltonian in Two Methods in Quantum Mechanics on Submanifold},
Prog. Theor. Phys. {\bf 87} (1992) 513--517.

\bibitem{fujii}
Takagi S., Tanzawa T.,
{\it Quantum Mechanics of a Particle Confined to a Twisted Ring}, 
Prog. Theor. Phys. {\bf 87}  (1992) 561--568;
Fujii K., Ogawa N.,
{\it Generalization of Geometry-Induced Gauge Structure to Any Dimensional Manifold}, 
Prog. Theor. Phys. {\bf 89} (1993) 575--578.

\end{thebibliography}
\end{document}